\documentclass[aps,prb,twocolumn,showpacs,reprint,longbibliography,final]{%
revtex4-1}
\usepackage{hyperref}
\usepackage{amsfonts}
\usepackage{amssymb}
\usepackage{amsmath}
\usepackage{graphicx}
\usepackage[utf8]{inputenc}
\usepackage{color}

\begin{document}

\title{Self-energy effects in functional renormalization group flows of the 
two-dimensional \texorpdfstring{$t$-$t'$}{t-t'} Hubbard model away from van 
Hove filling}

\date{\today}

\author{Andreas Eberlein}

 \affiliation{Department of Physics, Harvard University, Cambridge, 
Massachusetts 02138, USA}
\affiliation{Max Planck Institute for Solid State Research, 
 D-70569 Stuttgart, Germany}

\pacs{05.10.Cc, 71.10.Fd}

\newcommand{\etal}{\textit{et~al.}}
\newcommand{\ie}{\textit{i.\,e.}}
\newcommand{\bs}{\boldsymbol}

\newcommand{\tr}{\operatorname{tr}}

\newcommand{\V}[1]{\Gamma^{(4)\Lambda}_{#1}}

\newcommand{\VPH}[1]{V^{\text{PH},\Lambda}_{#1}}
\newcommand{\VPP}[1]{V^{\text{PP},\Lambda}_{#1}}

\newcommand{\dVPHdL}[1]{\dot V^{\text{PH},\Lambda}_{#1}}
\newcommand{\dVPPdL}[1]{\dot V^{\text{PP},\Lambda}_{#1}}

\newcommand{\G}[1]{G^\Lambda_{#1}}
\newcommand{\SL}[1]{S^\Lambda_{#1}}

\newcommand{\mtau}[2]{\tau^{(#1)}_{#2}}

\newcommand{\intdrei}[1]{\int\negthickspace\negthinspace\frac{d^3 #1}{(2\pi)^3}}


\hyphenation{counter-term}

\begin{abstract}
We study the impact of the fermionic self-energy on one-loop functional 
renormalization group flows of the two-dimensional $t$-$t'$ Hubbard model, with 
emphasis on electronic densities away from van Hove filling. In the presence of 
antiferromagnetic hot spots, antiferromagnetic fluctuations lead to a flattening 
of the Fermi surface, shift magnetic phase boundaries and significantly enhance 
critical scales. We trace back this effect to the presence of a magnetic first 
order transition. For some parameters, the first order character of the latter 
is reduced by self-energy effects. For reliably determining phase diagrams, the 
fermionic self-energy should be taken into account in functional renormalization 
group studies if scattering between hot spots is important.
\end{abstract}

\maketitle

\section{Introduction}
\label{sec:Introduction}
The Hubbard model is a paradigmatic model system for the description of 
correlated electrons in solids. Despite the simplicity of the model, an exact 
solution is only available in one spatial dimension. The two-dimensional model 
is believed to contain the essential ingredients to describe high-temperature 
superconductivity in cuprates. Similar to these materials, the model shows 
antiferromagnetism in its ground state at half-filling and becomes 
superconducting when sufficiently doped away from 
half-filling~\cite{Scalapino2012}. The lightly doped regime of cuprate 
materials and the Hubbard model are, however, relatively poorly understood. 
Achieving an accurate description of the properties of the Hubbard model in this 
regime is therefore highly desirable. 

Recent progress in the development of some numerical methods allowed to 
achieve agreement between the results in certain parameter regimes, but 
discrepancies remain for the important intermediate coupling regime at small 
doping~\cite{LeBlanc2015}. In this regime, several different states have 
very similar energies and an accurate determination of the ground state is 
difficult. One source of difficulty is the limitation of many numerical methods 
to relatively small systems. It therefore seems useful to complement numerical 
studies of the model with more analytical approaches that have access to 
low-energy Fermi surface instabilities in the thermodynamic limit. One such 
method is the functional renormalization 
group~\cite{Wetterich1993,Salmhofer2001,Metzner2012} (fRG).

This method treats all interaction channels on equal footing and is thus well 
suited for the study of competing order. One of its major successes was 
the unbiased detection of $d$-wave superconductivity in the ground state of the 
two-dimensional Hubbard model at weak 
coupling~\cite{Metzner2012,Friederich2011,Eberlein2014}. Results at weak 
coupling cannot be directly transferred to the cuprates, but allow to understand 
some qualitative features of their phase diagram in a controlled way. 
Until recently, the applicability of available truncation schemes of the fRG 
flow equations was limited to weak coupling, but this restriction may be 
removed by using non-trivial starting points for the fRG 
flow~\cite{Machado2010,Taranto2014,Wentzell2015}.

Most fRG studies for itinerant fermionic systems used a one-loop truncation, in 
which the fermionic self-energy and the two-particle vertex are 
renormalized~\cite{Katanin2004a,Metzner2012}. Motivated by the fact that the
information on most continuous phase transitions is encoded in the momentum 
dependence of the vertex, which diverges with the correlation length for certain 
combinations of momenta, the self-energy was usually neglected. A few studies 
investigated the influence of the self-energy on the flow and focused on the 
parameter region around van Hove filling. A general tendency towards a 
flattening of the Fermi surface was found in an early study slightly above van 
Hove filling, which however neglected the feedback of the self-energy on the 
flow~\cite{Honerkamp2001a}. More recent studies included the self-energy 
feedback on the flow, but were mostly restricted to a small region around van 
Hove filling, where interesting effects like competing instabilities or 
non-Fermi liquid behavior may already occur at weak 
coupling~\cite{Igoshev2011,Husemann2012,Giering2012}. In this parameter regime, 
it was found that renormalizing the Fermi surface has only a very small impact 
on the flow.

Despite the fact that deformations of the Fermi surface are important
perturbations in two-dimensional systems, their impact on functional 
renormalization group flows away from van Hove filling has not been fully 
addressed, yet. Knowing the impact of the self-energy on fRG flows in a broader 
parameter regime is certainly useful to judge the reliability of flows with 
perturbative as well as non-trivial starting points. Moreover, the one-loop fRG 
was also applied to model systems for the description of unconventional 
superconductivity in pnictide and ruthenate materials~\cite{Platt2013} and used 
to provide an explanation for subtle features in the momentum dependence of the 
superconducting gap. As these works neglected the fermionic self-energy, it 
would be interesting to know how robust their results are if the self-energy 
were taken into account.

In this work, we study the impact of the self-energy on one-loop fRG flows away 
from van Hove filling and at zero temperature. As in former studies, flows at 
van Hove filling are not changed qualitatively when the Fermi surface is 
renormalized. A similar conclusion holds for electron fillings below van Hove 
filling. In the presence of antiferromagnetic hot spots, \ie\ intersection 
points of the Fermi surface and the boundary of the magnetic Brillouin zone, 
the renormalization of the Fermi surface via the self-energy has a sizable 
quantitative impact on critical scales and shifts the boundaries between 
regimes with different leading instabilities. This is caused by a flattening of 
the Fermi surface, which leads to improved nesting and enhances 
antiferromagnetic fluctuations, thereby enlarging the parameter regime with 
antiferromagnetism as leading instability. Combining fRG flows with a mean-field 
(MF) analysis~\cite{Reiss2007,JWang2014} for the magnetic phase diagram, we 
trace back the large impact of the self-energy on the flow to the presence of a 
magnetic first-order phase transition.

This paper is organized as follows. Section~\ref{sec:ModelMethod} briefly 
introduces the Hubbard model and the fRG. Section~\ref{sec:Results} describes 
results mostly from one-loop flows in static approximation, \ie\ where the 
frequency dependence of the vertex and the self-energy are neglected, as well as 
results from a combination of fRG and mean-field theory. A few results from a 
dynamic approximation are also discussed. Section~\ref{sec:Conclusion} contains 
a summary and conclusions.

\section{Model and Method}
\label{sec:ModelMethod}

\subsection{Model}
\label{sec:Method}
The Hubbard model describes spin-$\tfrac{1}{2}$ fermions with a local repulsive 
interaction on a lattice. Its Hamiltonian in second-quantization notation is 
given by
\begin{equation}
	H = \sum_{i,j,\sigma} t_{ij} c^\dagger_{i\sigma} c_{j\sigma} + 
			U \sum_i n_{i\uparrow} n_{i\downarrow},
\end{equation}
where $c^{(\dagger)}_{i\sigma}$ are annihilation (creation) operators for 
fermions with spin orientation $\sigma = \uparrow, \downarrow$ on lattice site 
$i$. We study this model on a two-dimensional square lattice and restrict the 
hopping of fermions $t_{ij}$ to nearest and next-nearest neighbor sites with 
amplitudes $-t$ and $-t'$, respectively. Fourier transformation of the hopping 
matrix yields the dispersion 
\begin{equation}   
	\epsilon(\bs k) = -2 t (\cos k_x + \cos k_y) - 4 t' \cos k_x 
\cos k_y.
\end{equation}
Fermions occupying the same lattice site interact via the local Coulomb 
interaction with strength $U$.

In the following, we set $t \equiv 1$ and use it as the unit of energy.

\subsection{Functional renormalization group}
\label{sec:fRG}
The functional renormalization group allows to resum perturbation theory in a 
scale-separated way and treats all interaction channels on an equal footing. 
Comprehensive introductions to the method can be found in 
Refs.~\onlinecite{Berges2002,Metzner2012,Platt2013,Kopietz2010}.

The starting point of the method is a functional flow equation for the 
effective action~\cite{Wetterich1993,Salmhofer2001}, the generating functional 
of one-particle irreducible (1PI) vertex functions. The functional flow equation 
is equivalent to an infinite hierarchy of flow equations for vertex functions. 
Truncating this hierarchy and formulating an ansatz for the effective action 
allows to derive a closed set of renormalization group equations for the 
latter. 

In this work, we employ a truncation at the two-particle level, in which 
self-energy feedback from the three-particle level is taken into 
account~\cite{Katanin2004a}. Assuming translational and spin rotation 
invariance, we formulate the ansatz
\begin{equation}
\begin{split}
	&\Gamma^\Lambda[\bar\psi,\psi] = \Gamma^{(0)\,\Lambda} + \sum_{k,\sigma} 
\Gamma^{(2)\,\Lambda}(k) \bar\psi_{k\sigma} \psi_{k\sigma} \\
&+ \frac{1}{4} \sum_{k_i,\sigma_i} 
\Gamma^{(4)\,\Lambda}_{\sigma_1 \sigma_2 \sigma_3 \sigma_4}(k_1, k_2, k_3, k_4) 
\bar\psi_{k_1 \sigma_1} \bar\psi_{k_2 \sigma_2} \psi_{k_3 \sigma_3} \psi_{k_4 
\sigma_4}
\end{split}
\end{equation}
for the effective action, where $k = (k_0, \bs k)$ combines Matsubara 
frequencies and momenta. Due to symmetries, the two-particle vertex 
$\Gamma^{(4)\,\Lambda}$ is non-zero only for $k_1 + k_2 = k_3 + k_4$ and 
$\sigma_1 = \sigma_4$, $\sigma_2 = \sigma_3$ or $\sigma_1 = 
\sigma_3$, $\sigma_2 = \sigma_4$. 

The regularized fermionic propagator $G^\Lambda(k) = 
-(\Gamma^{(2)\,\Lambda})^{-1}(k)$ is related to the regularized bare propagator 
$G_0^\Lambda(k)$ and the self-energy $\Sigma^\Lambda(k)$ via a Dyson equation,
\begin{equation}
	(G^\Lambda)^{-1}(k) = (G_0^\Lambda)^{-1}(k) - \Sigma^\Lambda(k). 
\end{equation}
The regularized bare propagator is given by
\begin{equation}
	(G^\Lambda_0)^{-1}(k) = i k_0 - \epsilon(\bs k) + \mu + R^\Lambda(k),
\end{equation}
where $\mu$ is the chemical potential and $R^\Lambda$ the regulator. We use 
an additive frequency regulator,
\begin{equation}
	R^\Lambda(k) = i \operatorname{sgn}(k_0) \sqrt{k_0^2 + \Lambda^2} - i k_0,
	\label{eq:Regulator}
\end{equation}
that replaces small frequencies $k_0$ by $\operatorname{sgn}(k_0) \Lambda$ in 
$(G^\Lambda_0)^{-1}$. This regulator has been used in several works 
before~\cite{Eberlein2013,Eberlein2014,Eberlein2014a,Maier2014} and it was 
found that critical scales for pairing instabilities provide a very good 
estimate for the maximum amplitude of the ground state pairing gap.

Within the truncation scheme proposed by Katanin~\cite{Katanin2004a}, we obtain 
flow equations for the self-energy,
\begin{equation}
	\frac{d}{d\Lambda}\Sigma^\Lambda(k) = \sum_\sigma \intdrei{p} 
\Gamma^{(4)\,\Lambda}_{\uparrow\sigma\sigma\uparrow}(k,p,p,k) S^\Lambda(p),
\label{eq:SelfEnergy}
\end{equation}
where $S^\Lambda(k) = \frac{d}{d\Lambda} G^\Lambda(k)|_{\Sigma^\Lambda = 
\operatorname{const.}}$ is the fermionic single-scale propagator, and the 
two-particle vertex
\begin{widetext}
\begin{equation}
	\frac{d}{d\Lambda}\Gamma^{(4)\,\Lambda}_{\sigma_1 \sigma_2 \sigma_3 
\sigma_4}(k_1, k_2, k_3, k_4) = \Pi^{\text{PH},d}_{\sigma_1 \sigma_2 \sigma_3 
\sigma_4}(k_1, k_2, k_3, k_4) - \Pi^{\text{PH},cr}_{\sigma_1 \sigma_2 \sigma_3 
\sigma_4}(k_1, k_2, k_3, k_4)  - \frac{1}{2} \Pi^\text{PP}_{\sigma_1 \sigma_2 
\sigma_3 \sigma_4}(k_1, k_2, k_3, k_4),
\end{equation}
where the contributions on the right hand side are the direct particle-hole, 
crossed particle-hole and particle-particle diagram, respectively. They are 
given by
\begin{equation}
\begin{split}
	\Pi^{\text{PH},d}_{\sigma_1 \sigma_2 \sigma_3 
\sigma_4}(k_1, k_2, k_3, k_4) &= \sum_{p,\sigma_1',\sigma_2'} 
\frac{d}{d\Lambda}\Bigl(G^\Lambda(p-\tfrac{q}{2}) 
G^\Lambda(p+\tfrac{q}{2})\Bigr) \Gamma^{(4)\,\Lambda}_{\sigma_1 \sigma_1' 
\sigma_2' \sigma_4}\bigl(k_1, p-\tfrac{q}{2}, p + \tfrac{q}{2}, k_4\bigr)\\
	&\hspace{5em}\times \Gamma^{(4)\,\Lambda}_{\sigma_2' \sigma_2 \sigma_3 
\sigma_1'}\bigl(p + \tfrac{q}{2}, k_2, k_3, p-\tfrac{q}{2}\bigr)|_{q = k_3 - 
k_2}\\
	&= \Pi^{\text{PH},cr}_{\sigma_2 \sigma_1 \sigma_3 \sigma_4}(k_2, k_1, k_3, 
k_4)
\end{split}
\end{equation}
\begin{equation}
\begin{split}
	\Pi^\text{PP}_{\sigma_1 \sigma_2 \sigma_3 \sigma_4}(k_1, k_2, k_3, k_4) &= 
\sum_{p,\sigma_1',\sigma_2'} 
\frac{d}{d\Lambda}\Bigl(G^\Lambda(\tfrac{q}{2}-p)G^\Lambda(\tfrac{q}{2}
+p)\Bigr) \Gamma^{(4)\,\Lambda}_{\sigma_1 \sigma_2 \sigma_2' 
\sigma_1'}(k_1,k_2,\tfrac{q}{2}-p,\tfrac{q}{2}+p)\\
	&\hspace{5em} \times \Gamma^{(4)\,\Lambda}_{\sigma_1'\sigma_2'\sigma_3 
\sigma_4}(\tfrac{q}{2}+p,\tfrac{q}{2}-p,k_3,k_4)|_{q = k_1 + k_2}.
\end{split}
\end{equation}
\end{widetext}
Within this truncation, on the right hand side of the flow equation the 
scale-derivative of the full propagator,
\begin{equation}
	\frac{d}{d\Lambda}G^\Lambda(k) = S^\Lambda(k) + G^\Lambda(k) 
\dot\Sigma^\Lambda(k) G^\Lambda(k),
\end{equation}
appears instead of the single-scale propagator. This leads to a better treatment 
of self-energy corrections~\cite{Katanin2004a}.

In order to make the flow equations amenable for a numerical solution, we 
decompose the vertex into interactions channels and derive flow equations for 
the effective interaction in each channel as described in 
Refs.~\onlinecite{Karrasch2008,Husemann2009}. Our ansatz and notation are very 
similar to those of Ref.~\onlinecite{Eberlein2014}. We write
\begin{equation}
\begin{split}
	\Gamma^{(4)\,\Lambda}_{\sigma_1 \sigma_2 \sigma_3 \sigma_4}&(k_1, k_2, 
k_3, k_4) = u_{\sigma_1 \sigma_2 \sigma_3 \sigma_4}(k_1, k_2, k_3, k_4)\\
	& + V^{\text{PH},\Lambda}_{\sigma_1 \sigma_2 \sigma_3 
\sigma_4}(\tfrac{k_1+k_4}{2}, \tfrac{k_2+k_3}{2}; k_3-k_2) \\
	& - V^{\text{PH},\Lambda}_{\sigma_2 \sigma_1 \sigma_3 
\sigma_4}(\tfrac{k_2+k_4}{2}, \tfrac{k_1+k_3}{2}; k_3-k_1) \\
	& + V^{\text{PP},\Lambda}_{\sigma_1 \sigma_2 \sigma_3 
\sigma_4}(\tfrac{k_1-k_2}{2}, \tfrac{k_4-k_3}{2}; k_1 + k_2).
\end{split}
\end{equation}
for the vertex, where $u$ is the antisymmetrized bare interaction, and
$V^{\text{PH},\Lambda}$ and $V^{\text{PP},\Lambda}$ describe fluctuation 
corrections in the particle-hole and particle-particle channels, respectively. 
The first two momentum arguments of the latter two functions are fermionic 
relative momenta while the third is a bosonic momentum transfer or total 
momentum. For the Hubbard model, $u$ reads
\begin{equation}
\begin{split}
	u_{\sigma_1 \sigma_2 \sigma_3 \sigma_4}(k_1, k_2, k_3, k_4) &= U 
\delta_{k_1+k_2,k_3+k_4}\\
	&\times (\delta_{\sigma_1\sigma_4} \delta_{\sigma_2 \sigma_3} 
- \delta_{\sigma_1 \sigma_3} \delta_{\sigma_2 \sigma_4}).
\end{split}
\end{equation}
For the fluctuation corrections in the particle-hole channel, we make the ansatz
\begin{equation}
\begin{split}
	V^{\text{PH},\Lambda}_{\sigma_1 \sigma_2 \sigma_3 \sigma_4}(k, k'; q) &= (2 
\delta_{\sigma_1 \sigma_3} \delta_{\sigma_2 \sigma_4} - 
\delta_{\sigma_1 \sigma_4} \delta_{\sigma_2 \sigma_3}) M^\Lambda_{k k'}(q) \\
 &\ + \delta_{\sigma_1 \sigma_4} \delta_{\sigma_2 \sigma_3} C^\Lambda_{k k'}(q),
\end{split}
\end{equation}
where $C^\Lambda$ and $M^\Lambda$ are a density-density interaction $\sim n n$ 
and a spin-spin interaction $\sim \bs s \cdot \bs s$, respectively. The ansatz 
for the fluctuation corrections in the particle-particle channel reads
\begin{equation}
\begin{split}
	V^{\text{PP},\Lambda}_{\sigma_1 \sigma_2 \sigma_3 \sigma_4}(k, k'; q) &= 
\delta_{\sigma_1 \sigma_4} \delta_{\sigma_2 \sigma_3} P^\Lambda_{k k'}(q) \\
	& - \delta_{\sigma_1 \sigma_3} \delta_{\sigma_2 \sigma_4} P^\Lambda_{k, 
-k'}(q).
\end{split}
\end{equation}
Inserting these ansatzes into the flow equation for the vertex and assigning 
diagrams to interaction channels according to the transfer momenta in the 
fermionic loop integrals, we obtain the flow equations
\begin{gather}
	\frac{d}{d\Lambda} P^\Lambda_{k k'}(q) = -\frac{1}{2} 
\Pi^{\text{PP},\Lambda}_{\uparrow\downarrow\downarrow\uparrow}(k+\tfrac{q}{2},
\tfrac{q}{2}-k,\tfrac{q}{2}-k',k'+\tfrac{q}{2}),\\
	\frac{d}{d\Lambda} M^\Lambda_{k k'}(q) = \frac{1}{2} 
\Pi^{\text{PH},d}_{\uparrow\downarrow\uparrow\downarrow}(k+\tfrac{q}{2}, 
k'-\tfrac{q}{2}, k'+\tfrac{q}{2}, k-\tfrac{q}{2}),\\
\begin{split}
\frac{d}{d\Lambda} C^\Lambda_{k k'}(q) &= 
\Pi^{\text{PH},d}_{\uparrow\downarrow\downarrow\uparrow}(k+\tfrac{q}{2}, 
k'-\tfrac{q}{2}, k'+\tfrac{q}{2}, k-\tfrac{q}{2})\\
	& + \frac{d}{d\Lambda}M^\Lambda_{k k'}(q).
\end{split}
\end{gather}
More detailed expressions for the flow equations can be found in 
Appendix~\ref{sec:app:FlowEq}.

\subsection{Approximation scheme for vertex and self-energy}
\label{sec:Approx}
In the following we describe the approximations and parametrizations for the 
vertex and the self-energy that are applied within the channel-decomposition 
scheme. The framework of approximations is very similar to that in 
Refs.~\onlinecite{Eberlein2013,Eberlein2014}. Differently from these works, we 
do not introduce order parameters and analyze the leading instabilities of the 
flow.

The fluctuation corrections in the particle-hole and particle-particle channels 
are described as boson-mediated interactions. Former fRG studies identified the 
$s$- and $d_{x^2-y^2}$-wave channels as those yielding the largest 
contributions to the flow in the parameter regime that is relevant in this 
work~\cite{Zanchi2000,Halboth2000b,Honerkamp2001a,Husemann2009,Giering2012}. 
For the pairing, magnetic and charge fluctuations, we make the ansatzes
\begin{gather}
	P^\Lambda_{k k'}(q) = P^\Lambda_s(q) + P^\Lambda_d(q) f_d(\bs k) f_d(\bs 
k'),\\
	M^\Lambda_{k k'}(q) = M^\Lambda_s(q) + M^\Lambda_d(q) f_d(\bs k) f_d(\bs 
k'),\\
	C^\Lambda_{k k'}(q) = C^\Lambda_s(q) + C^\Lambda_d(q) f_d(\bs k) f_d(\bs k'),
\end{gather}
where $f_d(\bs k) = \cos k_x - \cos k_y$ is a lattice form factor with $d$-wave 
symmetry. The exchange propagators $P^\Lambda_i$, $M^\Lambda_i$ and 
$C^\Lambda_i$ describe mediated 
interactions in the $s$- and $d$-wave channels. The second contribution in the 
charge channel captures density-wave fluctuations with a $d$-wave form 
factor~\cite{Halboth2000b,Husemann2012b,Sachdev2013,Meier2014}. 

In this work we discuss two approximation schemes: A dynamic and a static 
approximation. In the dynamic approximation, we keep the dependence of the 
exchange propagators on the bosonic frequency $q_0$ and also the frequency 
dependence of the self-energy. In the static approximation, we neglect all 
frequency dependences and evaluate the exchange propagators at $q_0 = 0$. More 
details on the description of the momentum and frequency dependence of exchange 
propagators and the numerical solution of the flow equations can be found in 
Appendix~\ref{sec:NumDetailsAppendix}.

The central theme of this paper is to study the impact of the fermionic 
self-energy on fRG flows away from van Hove filling. We thus like to keep track 
of interaction-induced deformations of the Fermi surface, but also of the 
renormalization of the fermionic quasiparticles. At low scales we expect the 
momentum dependence of the self-energy parallel to the Fermi surface to be more 
important than the dependence perpendicular to it, and thus neglect the latter. 
We subdivide the Brillouin zone into patches (as in the $N$-patch approximation 
for the vertex~\cite{Zanchi2000}) and evaluate the self-energy 
$\Sigma^\Lambda(k)$ in each patch for the Fermi momentum in the middle of the 
patch. The frequency dependence is discretized on a non-equidistant grid with 
higher density of grid points at low frequencies, as for the exchange 
propagators. This reduces the self-energy in the dynamic approximation to a 
two-dimensional function of frequency and angle along the Fermi surface. At 
intermediate points, the self-energy is determined by linear interpolation. In 
the static approximation, the one-loop flow does not generate a frequency 
dependence of the self-energy and we evaluate it at $k_0 = 0$. The same 
approximations as for $\Sigma^\Lambda(k)$ are used for $\frac{d}{d\Lambda} 
\Sigma^\Lambda(k)$ appearing in the scale-differentiated propagator on the 
right hand side of the flow equations. In Sec.~\ref{sec:Results}, we 
compare results from fRG flows using these approximations for the self-energy 
with results from fRG flows using different approximations like 
neglecting the self-energy completely. The latter approximation is widely used 
in the literature.

\section{Results}
\label{sec:Results}
In this section, we present results from a numerical solution of the flow 
equations for the ground state. We discuss the dependence of critical 
scales $\Lambda_c$ on the next-nearest neighbor hopping $t'$, the fermionic 
density $n$ and various approximations for the self-energy $\Sigma$. We 
present results for a moderate interaction strength $U = 3$. The critical scale 
$\Lambda_c$ is defined as the scale $\Lambda$ where the largest component of 
exchange propagators exceeds $50 t$. When extrapolating the flow, it would 
diverge at slightly lower scales. All flows were evaluated in the presence of a 
fixed chemical potential $\mu$ and the fermionic density was determined from the 
full fermionic propagator at the critical scale.

In the following, we first discuss results for critical scales from static 
one-loop flows. In order to reduce the impact of the choice of the regulator on 
our conclusions, we compare the established trends with mean-field calculations 
that take the fRG results as input. At the end of this section we briefly 
discuss results from a dynamic one-loop approximation where the vertex and the 
self-energy also depend on frequency.

\subsection{Static one-loop flows}
\begin{figure}
	\centering
	\includegraphics[width=0.9\linewidth]{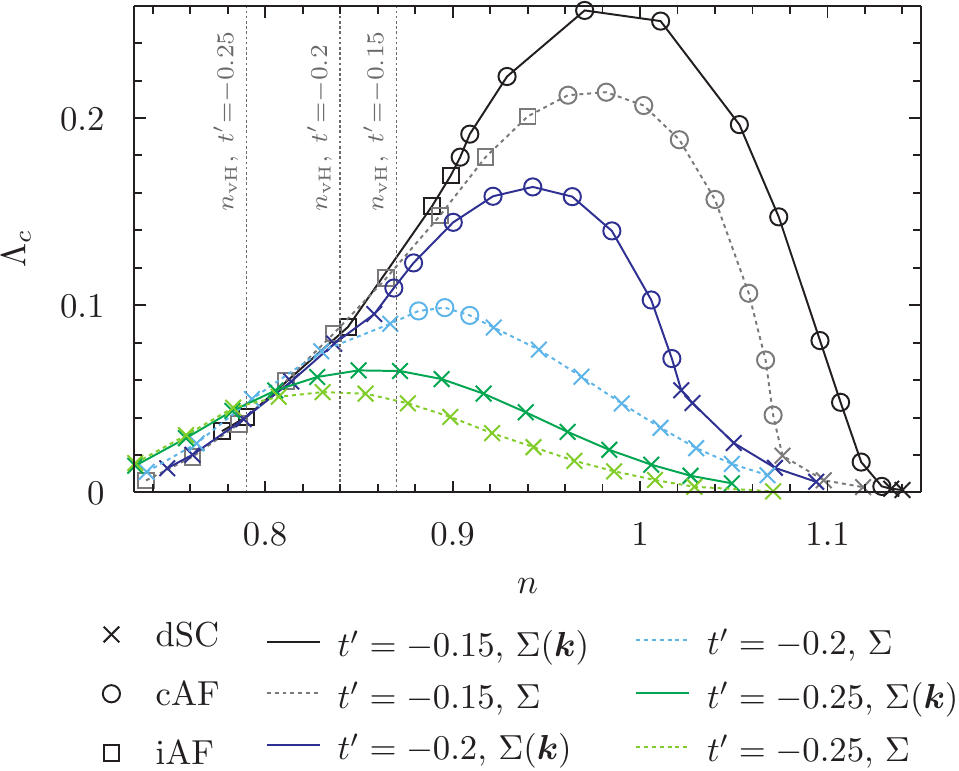}
	\caption{(Color online) Critical scales $\Lambda_c$ of one-loop fRG flows in 
static approximation for $U = 3$ and different values of $t'$. For each value of 
$t'$, critical scales from flows with momentum-dependent self-energy 
$\Sigma^\Lambda(\boldsymbol k)$ (labeled ``$\Sigma(\bs k)$'', full lines) and 
with momentum-independent self-energy $\Sigma^\Lambda(\boldsymbol k) = 
\Sigma^\Lambda$ (labeled ``$\Sigma$'', dashed lines) are compared. Symbols 
represent the leading instability (crosses for d-wave superconductivity (dSC), 
circles for commensurate antiferromagnetism (cAF) and squares for 
incommensurate antiferromagnetism(iAF)). The vertical dashed lines mark van 
Hove filling for the different values of $t'$.}
	\label{fig1}
\end{figure}
In the following, we present results from static one-loop flows. In 
Fig.~\ref{fig1} we compare critical scales $\Lambda_c$ as obtained from flows 
with momentum-dependent ($\Sigma^\Lambda(\bs k)$) or momentum-independent 
($\Sigma^\Lambda(\bs k) = \Sigma^\Lambda$ computed as Fermi surface average) 
self-energy for $U = 3$ and different values of $t'$. As discussed below 
(see Fig.~\ref{fig6}), the latter results are very close to those obtained from 
flows where the self-energy is neglected completely. The filling where 
the saddle points of the fermionic dispersion are part of the Fermi surface 
(van Hove filling) plays an important role in deciding how large the impact of 
a momentum-dependent renormalization of the self-energy is. We define van Hove 
filling as the filling where the saddle points are part of the renormalized 
Fermi surface at the critical scale, as in Ref.~\onlinecite{Giering2012}. This 
is the case for $n_\text{vH} = 0.87$ for $t' = -0.15$, $n_\text{vH} = 0.83$ for 
$t' = -0.2$ and $n_\text{vH} = 0.79$ for $t' = -0.25$. For electron densities 
near van Hove filling, the renormalization of the self-energy practically does 
not influence the critical scale, in agreement with the 
literature~\cite{Giering2012}. This is different closer to half-filling, where 
antiferromagnetic hot spots exist. When taking the momentum dependence of the 
self-energy into account, the parameter region with a leading instability 
towards antiferromagnetism is enlarged and the critical scale significantly 
enhanced for the smaller values of $-t'$. For $t' = -0.25$, no magnetic 
instability is found in both approximations, but antiferromagnetic fluctuations 
and the critical scale for $d$-wave pairing are also enhanced.

\begin{figure}
	\centering
	\includegraphics[width=0.9\linewidth]{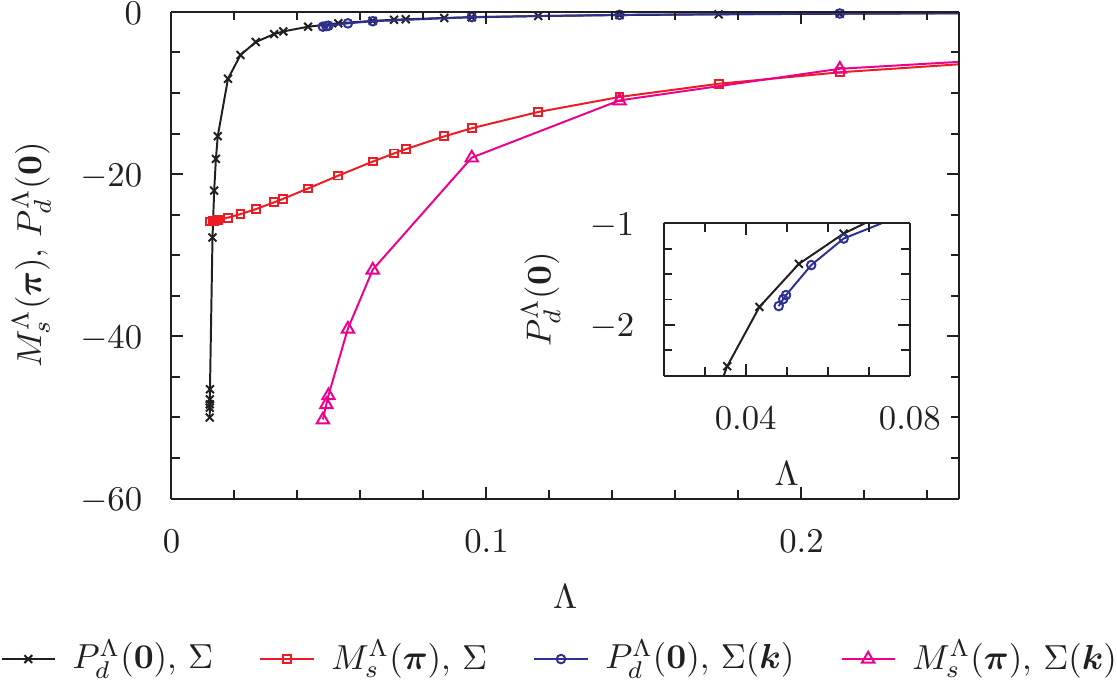}
	\caption{(Color online) Flow of exchange propagators in the magnetic channel 
$M^\Lambda_s(\boldsymbol q = \boldsymbol\pi)$ and in the $d$-wave pairing 
channel $P^\Lambda_d(\bs q = \boldsymbol 0)$ for $U = 3$, $t' = -0.15$ and $n = 
1.1$ as computed with renormalization of the momentum dependence of the 
self-energy $\Sigma^\Lambda(\bs k)$ (labeled ``$\Sigma(\bs k)$'') and by 
approximating the self-energy as momentum independent (labeled ``$\Sigma$''). 
The pairing interaction close to the critical scale of the flow with 
momentum-dependent self-energy is shown in the inset.}
	\label{fig2}
\end{figure}
On the electron-doped side, the leading instability can change from $d$-wave 
superconductivity to antiferromagnetism when renormalizing the Fermi surface. An 
exemplary flow of some couplings for such a case is shown in Fig.~\ref{fig2}. 
Using the momentum-independent approximation for the self-energy, the effective 
interaction in the magnetic channel saturates towards low scales and the 
$d$-wave pairing interaction eventually grows very strongly. Taking the momentum 
dependence of $\Sigma^\Lambda$ into account leads to a strong enhancement of 
antiferromagnetic fluctuations and eventually to a magnetic instability. At the 
critical scale of the latter flow, the $d$-wave pairing interaction is also 
enhanced by roughly $20\%$. For these parameters, the Fermi surface at the hot 
spots is not perfectly nested and the particle-hole bubble with transfer 
momentum $\bs q = \bs \pi$ thus finite, so that driving an antiferromagnetic 
instability requires a minimal coupling strength. In the flow with 
momentum-independent self-energy, the bare $U$ is effectively reduced by 
fluctuations below this threshold. In the momentum-dependent case, the 
renormalization of the Fermi surface leads to improved nesting around the hot 
spots, which in turn enhances antiferromagnetic fluctuations and gives rise to a 
magnetic instability.

\begin{figure}
	\centering
	\includegraphics[width=0.75\linewidth]{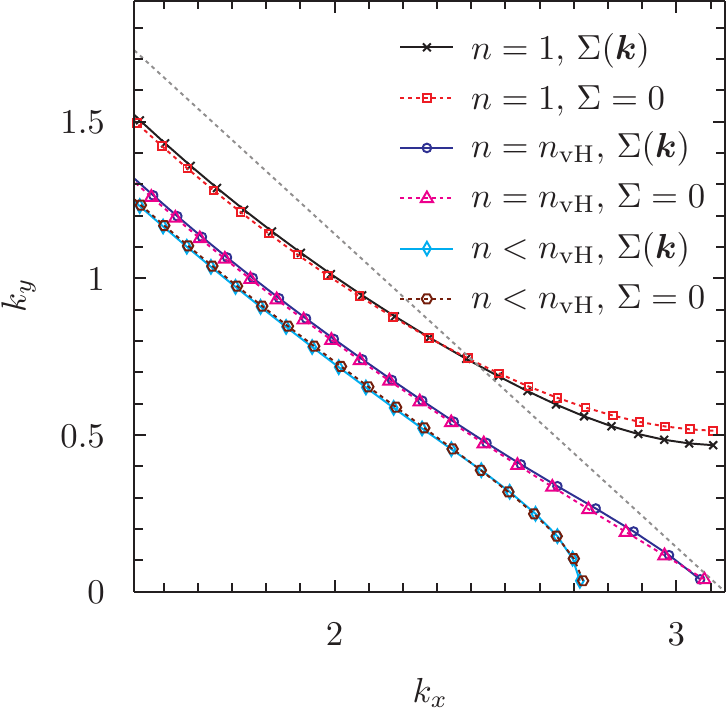}
	\caption{(Color online) Renormalized and bare Fermi surfaces from static 
one-loop fRG flows at the critical scale $\Lambda_c$ for $U = 3$ and 
$t' = -0.2$ for different electron fillings below van Hove filling ($n = 0.76$, 
$\Lambda_c = 0.02$), at van Hove filling ($n = n_\text{vH} = 0.83$, $\Lambda_c 
= 0.079$) and at half-filling ($n = 1$, $\Lambda_c = 0.103$). Renormalized 
(bare) Fermi surfaces are shown with full (dashed) lines and are labeled 
``$\Sigma(\bs k)$'' (``$\Sigma = 0$'').}
	\label{fig3}
\end{figure}
The improvement of nesting is mainly caused by antiferromagnetic 
fluctuations and can be seen in Fig.~\ref{fig3}. In 
this figure, we compare the bare and renormalized Fermi surfaces for $U = 3$, 
$t' = -0.2$ and different fermionic densities. The deformation of the Fermi 
surface for the parameters used in Fig.~\ref{fig2} is qualitatively very similar 
to that for $n = 1$ in Fig.~\ref{fig3}. At half-filling and van Hove filling ($n 
= n_\text{vH}$), the Fermi surface is flattened around the hot spots or the 
saddle points. A flattening of the Fermi surface around hot spots due to 
strong antiferromagnetic fluctuations was also observed before in the Hubbard 
model~\cite{Neumayr2003,Honerkamp2001a} and close to a spin-density wave quantum 
critical point with ordering wave vector $(\pi,\pi)$ in the spin-fermion 
model~\cite{Abanov2000,Metlitski2010b}.

\begin{figure}
	\centering
	\includegraphics[width=0.9\linewidth]{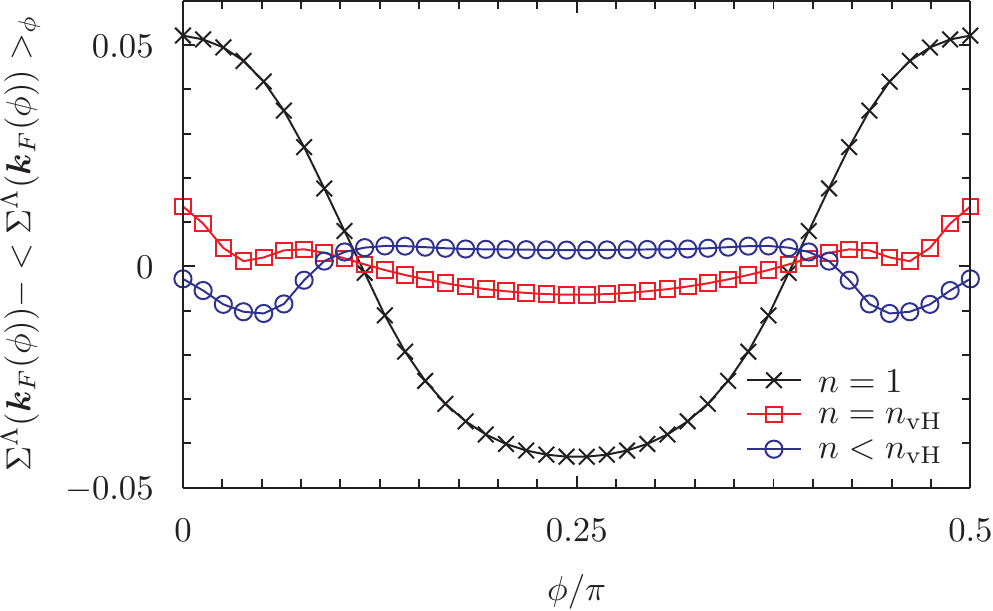}
	\caption{(Color online) Momentum dependence of the self-energy 
$\Sigma^\Lambda(\boldsymbol k)$ along the Fermi surface in the first quadrant 
of the Brillouin zone at the end of the flow for $U = 3$, $t' = -0.2$ and 
different fillings. The latter and the corresponding critical scales are the 
same as in Fig.~\ref{fig3}. In order to emphasize the momentum dependence, we 
subtracted the average of $\Sigma^\Lambda(\boldsymbol k)$ over all patches. 
$\phi = 0$ and $\phi = \frac{\pi}{2}$ correspond to the antinodal direction.}
	\label{fig4}
\end{figure}
The deformation of the Fermi surface as shown in Fig.~\ref{fig3} is caused by 
the self-energies shown in Fig.~\ref{fig4}. This figure shows 
$\Sigma^\Lambda(\bs k)$ along the Fermi surface after subtracting the average 
over all patches in order to highlight the momentum dependence. As expected from 
the small change of critical scales and Fermi surfaces (see Fig.~\ref{fig1} 
and~\ref{fig3}), at the lower fillings the magnitude of $\Sigma^\Lambda(\bs k)$ 
along the Fermi surface is very small. At half-filling, the self-energy is 
significantly larger and has a more pronounced momentum dependence. 

\begin{figure}
	\centering
	\includegraphics[width=0.9\linewidth]{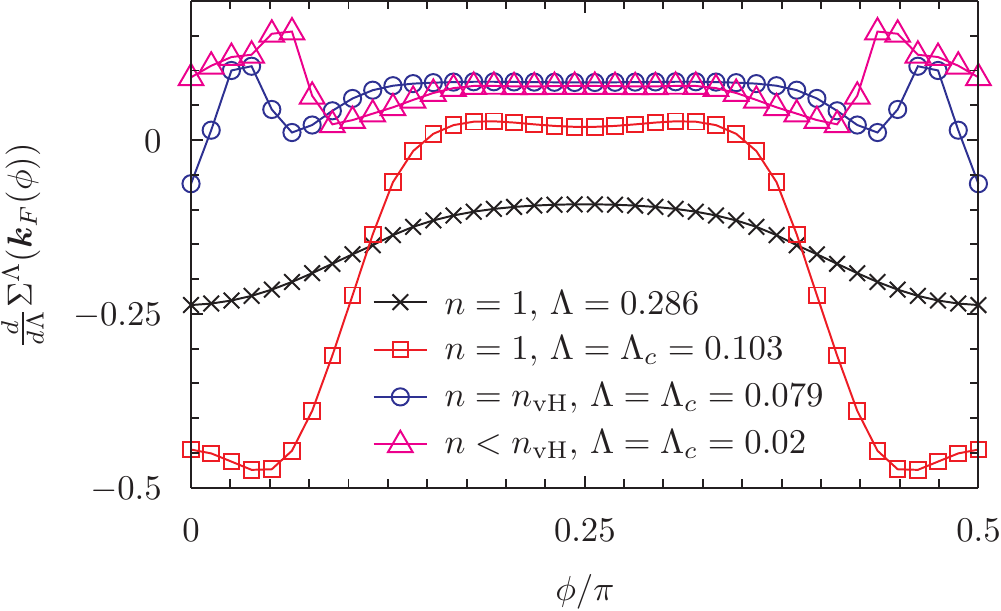}
	\caption{(Color online) Momentum dependence of the scale-derivative of the 
self-energy $\frac{d}{d\Lambda}\Sigma^\Lambda(\boldsymbol k)$ along the Fermi 
surface in the first quadrant of the Brillouin zone at the end of the flow for 
$U = 3$, $t' = -0.2$ and different fillings. The latter and the corresponding 
critical scales are the same as in Fig.~\ref{fig3}.}
	\label{fig5}
\end{figure}

\begin{figure}
	\centering
	\includegraphics[width=0.9\linewidth]{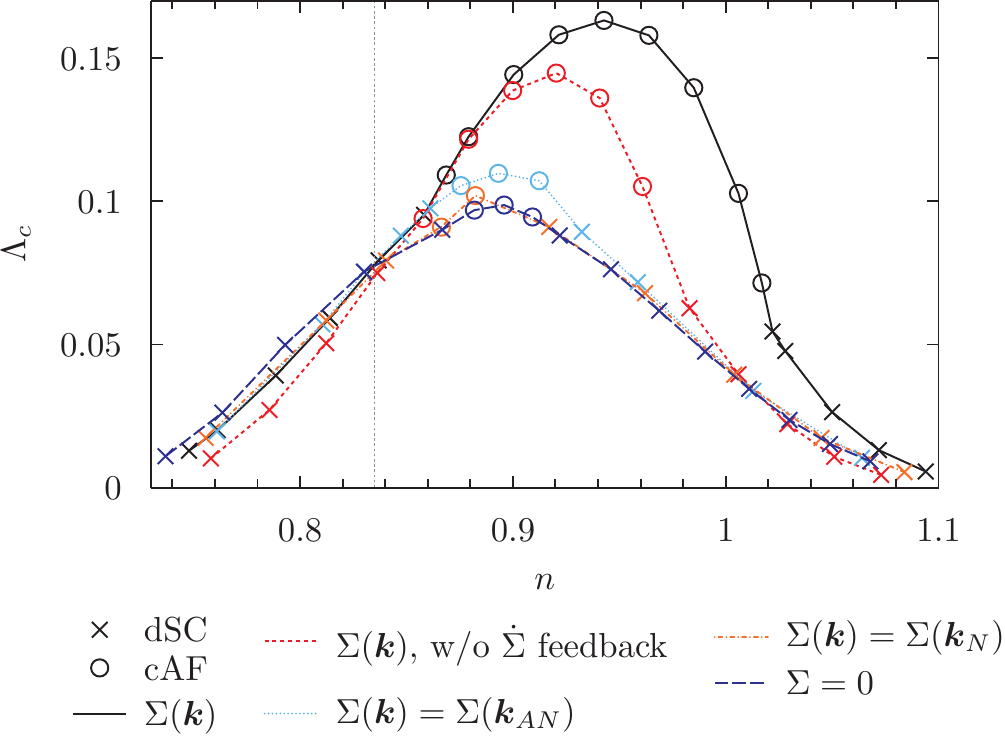}
	\caption{(Color online) Impact of different approximations for the 
self-energy $\Sigma^\Lambda(\bs k)$ on the critical scale $\Lambda_c$ of fRG 
flows in static approximation for $U = 3$ and $t' = -0.2$. Symbols represent the 
leading instability as in Fig.~\ref{fig1}. The different approximations are 
explained in the text.}
	\label{fig6}
\end{figure}

Parametrizing this self-energy in terms of renormalized hopping amplitudes to 
neighboring lattice sites as in Ref.~\onlinecite{Giering2012} should yield a 
very good approximation for all fillings. However, in the one-loop fRG 
truncation with self-energy feedback~\cite{Katanin2004a}, the scale-derivative 
of the self-energy also contributes on the right-hand side of the flow 
equation. As can be seen in Fig.~\ref{fig5}, 
$\frac{d}{d\Lambda}\Sigma^\Lambda(\bs k)$ for the half-filled system develops a 
strong dependence on momentum at low scales. A parametrization with a small 
number of renormalized hopping amplitudes may lead to an underestimation of 
$\frac{d}{d\Lambda}\Sigma^\Lambda(\bs k)$ and thus of the impact of 
fluctuations on the flow in this regime, in particular if the renormalization 
contributions to the hoppings are determined using Brillouin zone averages. 
Note that at lower fermionic densities 
(close to van Hove filling or below), $\frac{d}{d\Lambda}\Sigma^\Lambda(\bs k)$ 
remains small and is less important at low scales.

The relevance of the momentum dependence of $\Sigma^\Lambda(\bs k)$ and 
$\frac{d}{d\Lambda}\Sigma^\Lambda(\bs k)$ is illustrated in Fig.~\ref{fig6}, 
which shows the critical scales of static one-loop flows using different 
approximations for the self-energy and its scale-derivative. We show results for 
$t' = -0.2$ in this figure in order to highlight the effect of different 
approximations, which is somewhat amplified because $U$ is close to the 
critical value below which no instability to antiferromagnetism appears in the 
phase diagram when neglecting the momentum dependence of $\Sigma^\Lambda$. 
However, the conclusions for other values of $t'$ are qualitatively similar. In 
the figure, we show results for the approximations mentioned above and in 
addition for one-loop flows where i) $\Sigma^\Lambda(\bs k)$ is considered but 
the feedback of $\frac{d}{d\Lambda}\Sigma^\Lambda(\bs k)$ to the right-hand side 
of the flow equation neglected (labeled ``$\Sigma(\bs k)$, w/o $\dot \Sigma$ 
feedback''), ii) a momentum-independent self-energy is used and evaluated for 
the Fermi momentum in the antinodal direction closest to the saddle points 
(labeled ``$\Sigma(\bs k) = \Sigma(\bs k_{AN})$'') and iii) the same as (ii) but 
evaluated for a Fermi momentum in the nodal direction (labeled ``$\Sigma(\bs k) 
= \Sigma(\bs k_{N})$''). Evaluation of the momentum-independent self-energy in 
the nodal direction yields critical scales that are almost equal to those that 
result when the self-energy is neglected completely or evaluated as a Fermi 
surface average. The reason is that in this case the self-energy flows only 
weakly at low scales because the nodal points are closer together than 
$(\pi,\pi)$ and antiferromagnetic fluctuations thus ineffective. Evaluation of 
the self-energy in the antinodal direction has a somewhat larger effect on 
critical scales close to van Hove filling. In this case the influence of 
antiferromagnetic fluctuations is larger and changes in $\Sigma^\Lambda$ 
strongly influence the density of states at the Fermi level. Sufficiently away 
from van Hove filling, antiferromagnetic fluctuations cease to be effective in 
renormalizing $\Sigma^\Lambda$ in this approximation and the self-energy 
becomes unimportant. Taking the momentum dependence of $\Sigma^\Lambda$ into 
account but neglecting the feedback of $\dot \Sigma^\Lambda$, the critical 
scale is significantly enhanced in a certain density range above van Hove 
filling, but smaller than with full self-energy feedback. It is interesting to 
note that not only the deformation of the Fermi surface matters, but that the 
feedback of $\dot\Sigma^\Lambda$ also has a sizable impact on critical scales.

\subsection{Static one-loop flows and fRG+MF}
\label{subsec:RGMF}
Former fRG studies, which mainly focused on the parameter regime around van 
Hove filling, found that the renormalization of the Fermi surface has a small 
impact on the flow. In the last section we found that taking the fermionic 
self-energy into account can strongly enhance critical scales close to 
half-filling. This could be due to an underlying physical mechanism or due to 
using a different regulator. It is known that critical scales of fRG flows 
depend to some extent on the employed regularization scheme. An extreme example 
are forward scattering driven instabilities with $\bs q = \bs 0$, which cannot 
be detected as instabilities of one-loop fRG flows at zero temperature when 
using a momentum cutoff~\cite{Halboth2000a}, but which show up when using 
frequency~\cite{Husemann2009} or temperature~\cite{Honerkamp2001b} cutoff 
schemes. It would be interesting to better understand the origin of the larger 
impact of the self-energy found in the last section.

For this purpose we use a combination of fRG and mean-field theory (MF) to 
compute order parameters based on input from fRG 
flows~\cite{Reiss2007,JWang2014}, as order parameters should have a weaker 
dependence on regularization schemes than critical scales. This approach was 
already applied to study the competition of superconductivity with 
commensurate~\cite{Reiss2007,JWang2014} and 
incommensurate~\cite{Yamase2015-arXiv} antiferromagnetism in the ground state 
of the two-dimensional Hubbard model. For superconducting ground states, the 
method yielded superconducting gaps in good agreement with results from 
one-loop fRG flows into the symmetry broken phase~\cite{Eberlein2014}.

The increase of critical scales discussed in the last section is a 
consequence of the interplay between Fermi surface deformation and 
antiferromagnetic fluctuations. We therefore restrict ourselves to the 
computation of the magnetic phase diagram and consider only gaps due to 
commensurate antiferromagnetism in the mean-field calculation. Near half-filling 
and on the electron-doped side this is justified because the antiferromagnetic 
gap is only weakly affected by coexisting superconducting 
order~\cite{Yamase2015-arXiv}. For simplicity we do not further renormalize the 
normal self-energy in the mean-field calculation. Changes of the 
antiferromagnetic gap therefore reflect changes of the vertex and the Fermi 
surface due to self-energy feedback during the fRG flow. We expect that 
renormalizing the Fermi surface in the mean-field part of the calculation would 
further increase antiferromagnetic ordering tendencies.

After stopping the fRG flow at the critical scale $\Lambda_c$, which we take 
as the mean-field scale $\Lambda_\text{MF}$, we extract the irreducible 
vertex $\tilde U_{\bs k \bs k'}$ in the antiferromagnetic channel with transfer 
momentum $\bs Q = (\pi,\pi)$ from the full vertex in this channel,
\begin{equation}
	U_{\bs k \bs k'} = \sum_{\sigma} \epsilon_\sigma 
\Gamma^{(4)\,\Lambda_\text{MF}}_{\uparrow 
\sigma \sigma \uparrow}(k + Q, k', k' + Q, k)|_{k_0 = k_0' = 0},
\end{equation}
where $Q = (0, \bs Q)$ and $\epsilon_\uparrow = 1$, $\epsilon_\downarrow = -1$, 
via a Bethe-Salpeter-like integral equation,
\begin{equation}
\begin{split}
	U_{\bs k \bs k'} &= \tilde U_{\bs k \bs k'} + \int_{p_0} \int_{\bs p} \tilde 
U_{\bs k \bs p} G^{\Lambda_\text{MF}}(p) G^{\Lambda_\text{MF}}(p + Q) U_{\bs p 
\bs k'},
\end{split}
\end{equation}
where $\int_{p_0}$ and $\int_{\bs p}$ are shorthands for 
$\int\negthinspace\frac{d p_0}{2\pi}$ and $\int\negthinspace\frac{d^2 
p}{(2\pi)^2}$, respectively, and $G^{\Lambda_\text{MF}}$ is the fermionic 
propagator including the self-energy at scale $\Lambda_\text{MF} = \Lambda_c$. 
The irreducible vertex is 
inserted into the antiferromagnetic gap equation
\begin{equation}
	A_{\bs k} = \frac{1}{2} \int_{\bs k'} \tilde 
U_{\bs k \bs k'} \langle m_{\bs k'} \rangle,
\end{equation}
where $m_{\bs k} = a^\dagger_{\bs k\uparrow} a_{\bs k + \bs Q\uparrow} - 
a^\dagger_{\bs k\downarrow} a_{\bs k + \bs Q\downarrow}$ is the staggered 
magnetization and $a^{(\dagger)}$ are fermionic annihilation (creation) 
operators. The expectation values are computed using the mean-field Hamiltonian
\begin{equation}
\begin{split}
	H_\text{MF} = &\int_{\bs k} \sum_\sigma \bigl(\epsilon(\bs k) + 
\Sigma^{\Lambda_c}(\bs k)\bigr) a^\dagger_{\bs k \sigma} a_{\bs k \sigma}\\
	&+ \int_{\bs k} A_{\bs k} (m_{\bs k} - \tfrac{1}{2} \langle m_{\bs k} 
\rangle).
\end{split}
\end{equation}
Such an effective Hamiltonian formulation is possible because we use a static 
approximation. Note that the self-consistency equations are solved at $\Lambda 
= 0$, \ie\ in the absence of a regulator.

\begin{figure}
	\centering
	\includegraphics[width=0.9\linewidth]{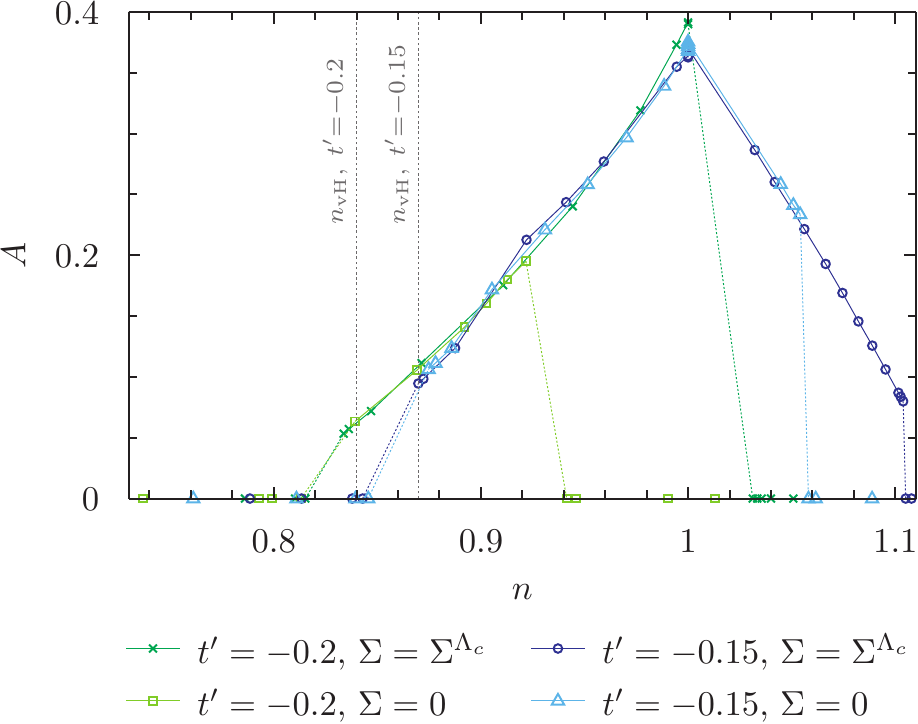}
	\caption{(Color online) Amplitude of the antiferromagnetic gap function for 
$U = 3$ and two values of $t'$. The results were obtained as solutions of the 
gap equation, based on fRG flows in which the normal self-energy was 
renormalized (labeled ``$\Sigma = \Sigma^{\Lambda_c}$'') or neglected completely 
(labeled ``$\Sigma = 0$''). The normal self-energy was not renormalized in the 
mean-field calculation. Dashed lines are guides to the eye at first order 
phase transitions. Dashed grey vertical lines mark van Hove filling.}
	\label{fig7}
\end{figure}
In Fig.~\ref{fig7} we show the magnetic phase diagram as obtained from solving 
the gap equation for $U = 3$ and two values of $t'$, comparing how the 
renormalization of the self-energy and its feedback in the fRG flow change the 
magnetic order parameter. The input for the gap equation is obtained (i) from 
a fRG flow in which the self-energy is neglected completely, $\Sigma = 0$, and 
(ii) from a fRG flow in which a momentum dependent self-energy $\Sigma(\bs k)$ 
is considered. In all cases, the antiferromagnetic order disappears 
via first-order transitions. The antiferromagnetic gap is almost unchanged where 
it appears in both approximations. For $t' = -0.2$ and $\Sigma = 0$, 
antiferromagnetism exists only close to van Hove filling. Renormalizing the 
normal self-energy in the fRG flow for this value of $t'$, the antiferromagnetic 
phase gets significantly larger and extends up to half-filling. For $t' = -0.15$ 
and $\Sigma = 0$, the antiferromagnetic phase already extends to the 
electron-doped side. Renormalizing the Fermi surface during the fRG flow leads 
to an extension of the antiferromagnetic phase and a strong reduction of the 
first-order transition between the antiferromagnetic and the paramagnetic metal. 
This shift of phase boundaries and the enlarged antiferromagnetic regimes are 
the reason for the increase of critical scales in the fRG flow as discussed in 
the last section. The reduction of the first-order character of the magnetic 
phase transition for $t' = -0.15$ seems to be generic for cases where the 
antiferromagnetic phase extends beyond half-filling. Note that the first order 
transitions to metallic states on the hole-doped side are artifacts of our 
approximation, as they would be preempted by first order transitions to metallic 
incommensurate antiferromagnetic states~\cite{Igoshev2010,Yamase2015-arXiv}. On 
the other hand our results suggest that self-energy corrections do not 
qualitatively modify the findings by Yamase~\etal~\cite{Yamase2015-arXiv} as 
self-energy corrections are of minor importance below and around van Hove 
filling.

\subsection{Dynamic one-loop flows}
\label{sec:dyn_flows}
In this section we briefly discuss what happens if the frequency dependence of 
the vertex and the self-energy are taken into account as described in 
Sec.~\ref{sec:Approx}. Our flow equations are very similar to those of 
Husemann~\etal~\cite{Husemann2012} and Giering and Salmhofer~\cite{Giering2012}. 
Here we also solve them away from van Hove filling in particular close to 
half-filling where antiferromagnetic hot spots exist on the Fermi surface.

\begin{figure}
	\centering
	\includegraphics[width=0.9\linewidth]{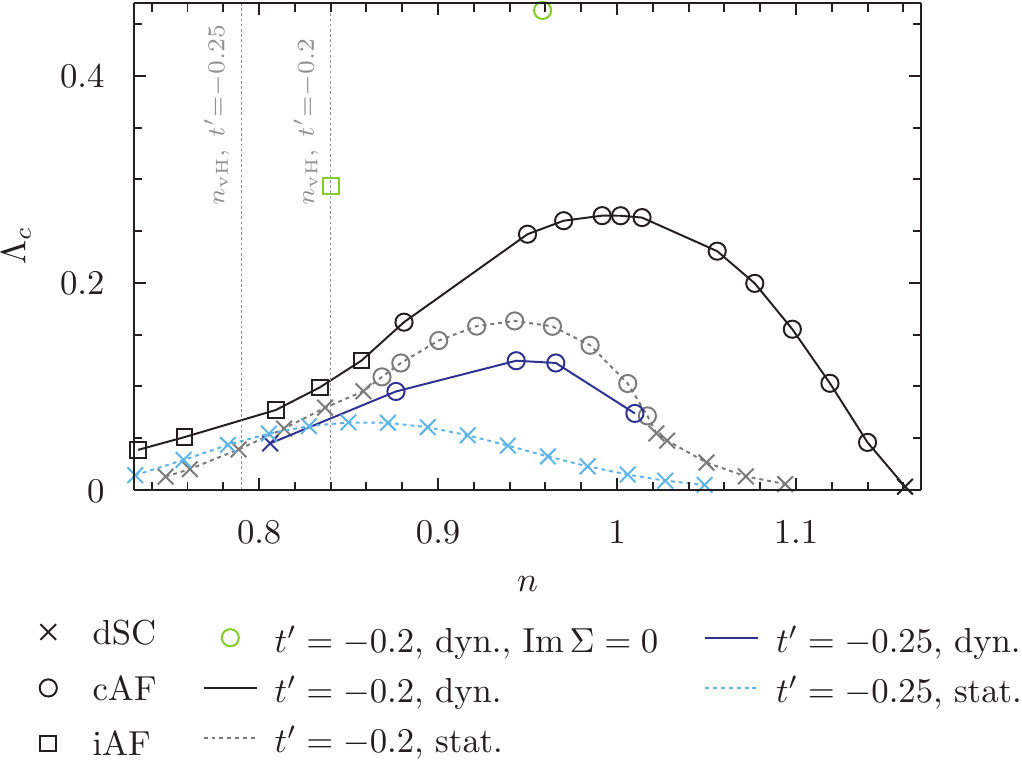}
	\caption{(Color online) Critical scales $\Lambda_c$ from dynamic 
(``dyn.'', full lines) and static (``stat.'', dashed lines) one-loop fRG flows 
for $U = 3$ and different values of $t'$. Symbols indicate the leading 
instability as in Fig.~\ref{fig1}. Also shown are a few data points from dynamic 
one-loop flows in which $\operatorname{Im}\Sigma^\Lambda$ was neglected 
(``$\operatorname{Im}\Sigma = 0$'', only symbols). The vertical dashed lines 
mark van Hove filling.}
	\label{fig8}
\end{figure}
We do not show results for the frequency dependence of the exchange propagators 
and the self-energy, as they are qualitatively similar to those in 
Refs.~\onlinecite{Husemann2012,Giering2012}. In Fig.~\ref{fig8}, we compare 
critical scales from static and dynamic one-loop flows for $U = 3$ and $t' = 
-0.2$ as well as $-0.25$. We observe that taking the frequency dependence of the 
self-energy and the vertex into account yields larger critical scales and a 
broader density range with antiferromagnetism as leading instability. For $t' = 
-0.25$, antiferromagnetic instabilities appear, which were not present in the 
static approximation. The increase of critical scales in comparison to the 
static approximation seems to be a peculiarity of the chosen regulator and 
should not be misunderstood in the sense that a frequency dependent vertex and 
self-energy, and in particular a reduced quasiparticle weight, enhance ordering 
tendencies. Fluctuation contributions are weighted differently in the static and 
the dynamic one-loop approximation. Which one yields smaller critical scales 
depends on the regularization scheme. For a smooth multiplicative frequency 
regulator~\cite{Husemann2012,Giering2012} and a sharp multiplicative momentum 
cutoff~\cite{Uebelacker2012}, it was found for the repulsive Hubbard model at 
van Hove filling that the critical scales were smaller in the dynamic than in 
the static approximation. For the attractive Hubbard model and the same 
regulator as employed in this work, the static approximation yielded smaller 
critical scales and superconducting gaps than the dynamic 
approximation~\cite{Eberlein2013}. Neglecting the frequency dependence of the 
self-energy (and thus the renormalization of the quasiparticle weight via 
$\operatorname{Im}\Sigma$), but taking the frequency dependence of the vertex 
into account, yields a further increase of critical scales for all regulators. 
This is shown in Fig.~\ref{fig8} for the regulator used in this work for two 
fillings. Note that although the critical scales are rather high in this case, 
they are still significantly smaller than the critical scales of RPA-like flows 
in which all fluctuation contributions are neglected (in the latter case we 
obtain $\Lambda_c = 0.58$ and $0.78$ for $n = 0.83$ and $1.07$, respectively).

\begin{figure}
	\centering
	\includegraphics[width=0.9\linewidth]{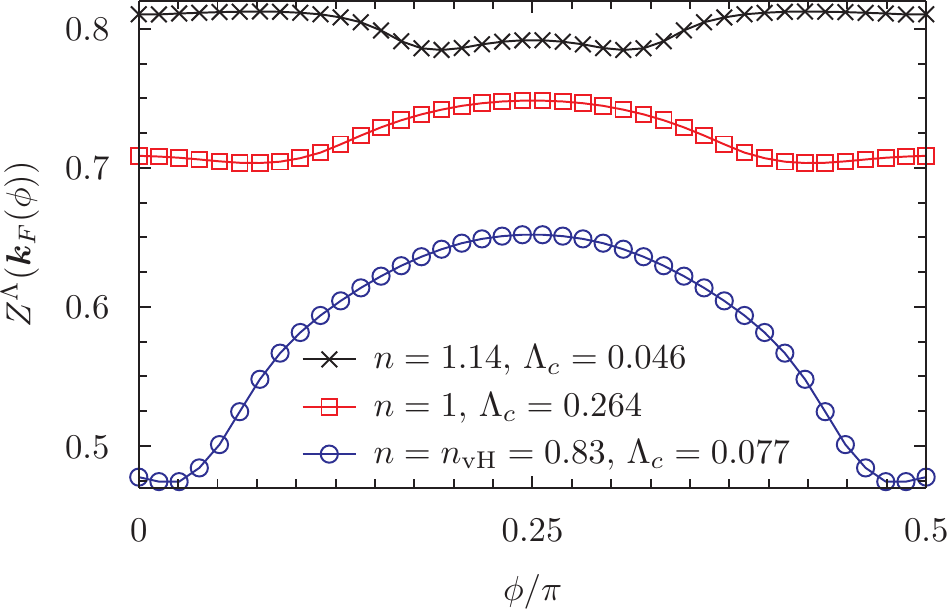}
	\caption{(Color online) Quasiparticle weight $Z^\Lambda$ along the Fermi 
surface at the end of the flow for $U = 3$, $t' = -0.2$ and various densities. 
$\phi = 0$ and $\pi/2$ denote the antinodal region.}
	\label{fig9}
\end{figure}

Computing the self-energy in the $N$-patch approximation allowed us to resolve 
its frequency dependence and its momentum dependence along the Fermi surface 
with high resolution. In Fig.~\ref{fig9}, we show the variation of the 
quasiparticle weight
\begin{equation}
	Z^\Lambda(\bs k) = \bigl(1 - \partial_{k_0} \operatorname{Im} 
\Sigma^\Lambda(k_0, \bs k)|_{k_0 = 0}\bigr)^{-1}
\end{equation}
along the Fermi surface for $U = 3$, $t' = -0.2$ and various fermionic 
densities. In all cases shown, the quasiparticle weight is smallest very 
close to the antiferromagnetic hot spots. At van Hove filling, the 
minimum is slightly shifted away from the saddle point of the fermionic 
dispersion because the antiferromagnetic fluctuations are incommensurate at low 
scales. The quasiparticle weight also shows a sizable anisotropy between the 
nodal and the antinodal direction, in agreement with former fRG studies for 
similar parameters~\cite{Honerkamp2003,Uebelacker2012}. With increasing filling, 
the anisotropy weakens and the minima shift towards the Brillouin zone diagonal. 
$Z^\Lambda(\bs k_F)$ being minimal at the hot spots is consistent with the 
behavior of the spin-fermion model at a spin-density wave quantum critical point 
with ordering momentum $\bs Q = (\pi,\pi)$~\cite{Lee2013}. It is interesting 
that the quasiparticle weight is largest for $n = 1.14$, although the critical 
scale is lower and the Fermi velocities at the hot spots are slightly more 
antiparallel than at half-filling.
\begin{figure}
	\centering
	\includegraphics[width=0.9\linewidth]{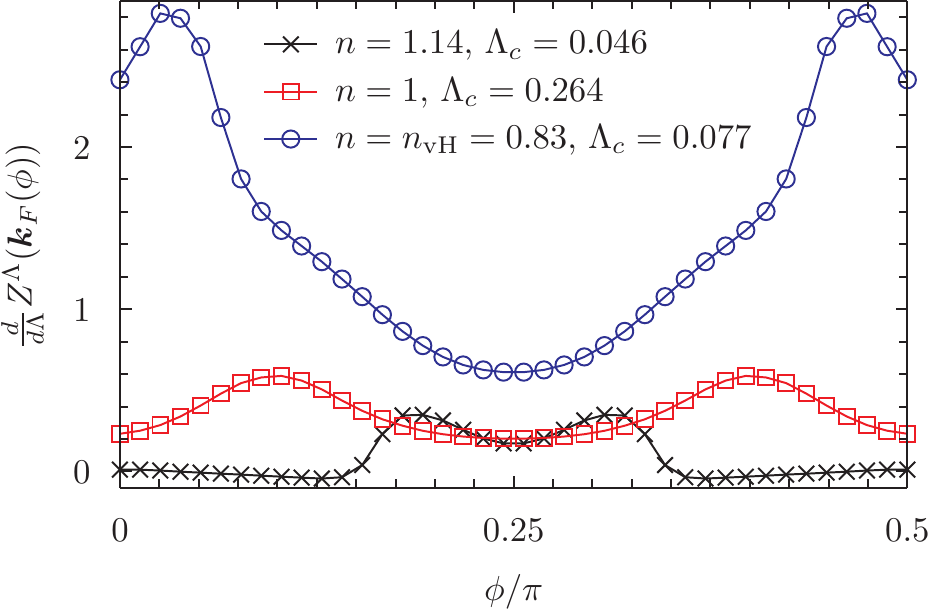}
	\caption{(Color online) Scale-derivative of the quasiparticle weight 
$\frac{d}{d\Lambda} Z^\Lambda$ along the Fermi surface at the end of the flow 
for $U = 3$, $t' = -0.2$ and the same densities as in Fig.~\ref{fig9}.}
	\label{fig10}
\end{figure}

The angular dependence of the scale-derivative of the quasiparticle weight is 
qualitatively similar to that of $Z^\Lambda$, as can be seen in 
Fig.~\ref{fig10}. In case the antiferromagnetic fluctuations are commensurate, 
the maximum is located at the antiferromagnetic hot spots. At van Hove filling, 
the maximum is shifted slightly away from the saddle points due to 
incommensurate magnetic fluctuations. Note that a positive 
$\frac{d}{d\Lambda}Z^\Lambda$ yields a reduction of $Z^\Lambda$ during the flow.

\section{Conclusion}
\label{sec:Conclusion}
In this paper, we have analyzed the impact of self-energy effects on functional 
renormalization group flows away from van Hove filling. For the latter filling, 
which was mostly studied before, our conclusions agree very well with the 
literature. In particular, fluctuation induced deformations of the Fermi surface 
have a small impact on the flow. This changes for higher fillings, where 
antiferromagnetic hot spots exist. In this regime, mostly antiferromagnetic 
fluctuations lead to a flattening of the Fermi surface, which itself amplifies 
the magnetic fluctuations. This effect is well known and here we showed that it 
significantly enhances ordering tendencies to commensurate antiferromagnetism 
and leads to higher critical scales. Interestingly, for the parameters 
considered in this work the maximum of critical scales is always shifted towards 
half-filling.

Using a combination of functional renormalization group and mean-field theory, 
we showed that self-energy corrections lead to strong shifts of magnetic phase 
boundaries and can reduce the first order character of antiferromagnetic phase 
transitions. The former is the underlying reason for the enhanced ordering 
tendencies and critical scales as found in the renormalization group flow.

Also considering the frequency dependence of the vertex and the self-energy 
within a dynamic approximation, we found a further increase of critical scales 
and enlargement of the parameter region with antiferromagnetic order. Note that 
the critical scales are nevertheless significantly smaller than in an RPA-type 
approximation where fluctuations are neglected.

On the question whether deformations of the Fermi surface have a qualitative 
impact on phase diagrams derived from one-loop functional renormalization group 
flows, our results suggest that they are indeed important close to first order 
phase transitions with finite ordering wave vectors as in the case of 
antiferromagnetism, or in the presence of hot spots. For lower electron 
densities below and around van Hove filling, they seem to be of minor 
importance.

\begin{acknowledgments}
I would like to thank S.~Andergassen, J.~Bauer, T.~Holder, W.~Metzner, A.~Toschi, 
K.~Veschgini, D.~Vilardi, J.~Wang and H.~Yamase for valuable discussions. This 
research was partially supported by the German National Academy of Sciences 
Leopoldina through grant LPDS~2014-13.
\end{acknowledgments}

\appendix

\begin{widetext}
\section{Flow equations for the vertex and the self-energy}
\label{sec:app:FlowEq}
In this appendix, we describe the flow equations for the vertex and the 
self-energy in somewhat more detail. Note that we exploit translation 
invariance, spin rotation invariance and time reversal symmetry for their 
derivation.

The flow equation for the effective interaction in the magnetic channel reads
\begin{equation}
	\frac{d}{d\Lambda} M^\Lambda_{k k'}(q) = \frac{1}{2} \intdrei{p} 
\frac{d}{d\Lambda}\bigl(G^\Lambda(p-\tfrac{q}{2}) 
G^\Lambda(p+\tfrac{q}{2})\bigr) 
\Gamma^{(4)\,\Lambda}_{\uparrow\downarrow\uparrow\downarrow}(k+\tfrac{q}{2}, 
p-\tfrac{q}{2}, p+\tfrac{q}{2}, k-\tfrac{q}{2}) 
\Gamma^{(4)\,\Lambda}_{\downarrow\uparrow\downarrow\uparrow}(k'-\tfrac{q}{2}, 
p+\tfrac{q}{2}, p-\tfrac{q}{2}, k'+\tfrac{q}{2}).
\end{equation}
Inserting the channel decomposition of the vertex, its component relevant for 
this flow equation reads
\begin{equation}
	\Gamma^{(4)\,\Lambda}_{\uparrow\downarrow\uparrow\downarrow}(k+\tfrac{q}{2}, 
p-\tfrac{q}{2}, p+\tfrac{q}{2}, k-\tfrac{q}{2}) = -U + 2 M^\Lambda_{k p}(q) + 
M^\Lambda_{\frac{k+p-q}{2}, \frac{k+p+q}{2}}(p-k) - C^\Lambda_{\frac{k+p-q}{2}, 
\frac{k+p+q}{2}}(p-k) - P^\Lambda_{\frac{k-p+q}{2}, \frac{p-k+q}{2}}(p+k),
\end{equation}
where 
$\Gamma^{(4)\,\Lambda}_{\downarrow\uparrow\downarrow\uparrow}(k'-\tfrac{q}{2}, 
p+\tfrac{q}{2}, p-\tfrac{q}{2}, k'+\tfrac{q}{2}) = 
\Gamma^{(4)\,\Lambda}_{\uparrow\downarrow\uparrow\downarrow}(k'+\tfrac{q}{2}, 
p-\tfrac{q}{2}, p+\tfrac{q}{2}, k'-\tfrac{q}{2})$ due to symmetries. For the 
effective interaction in the charge channel, we obtain
\begin{equation}
\begin{split}
	\frac{d}{d\Lambda}C^\Lambda_{k k'}(q) = \intdrei{p} &
\frac{d}{d\Lambda}\bigl(G^\Lambda(p-\tfrac{q}{2}) 
G^\Lambda(p+\tfrac{q}{2})\bigr) 
\Bigl(\Gamma^{(4)\,\Lambda}_{\uparrow\uparrow\uparrow\uparrow}(k+\tfrac{q}{2},
p-\tfrac{q}{2}, p+\tfrac{q}{2}, k - \tfrac{q}{2}) 
\Gamma^{(4)\,\Lambda}_{\downarrow\uparrow\uparrow\downarrow}(k'-\tfrac{q}{2}, p 
+ \tfrac{q}{2}, p - \tfrac{q}{2}, k'+\tfrac{q}{2}) \\
	& + 
\Gamma^{(4)\,\Lambda}_{\uparrow\downarrow\downarrow\uparrow}(k+\tfrac{q}{2}, 
p-\tfrac{q}{2}, p+\tfrac{q}{2}, k-\tfrac{q}{2}) 
\Gamma^{(4)\,\Lambda}_{\downarrow\downarrow\downarrow\downarrow}(k'-\tfrac{q}{2} 
, p+\tfrac{q}{2}, p-\tfrac{q}{2}, k'+\tfrac{q}{2})\Bigr) + \frac{d}{d\Lambda} 
M^\Lambda_{k k'}(q),
\end{split}
\end{equation}
where
\begin{gather}
\begin{split}
\Gamma^{(4)\,\Lambda}_{\uparrow\uparrow\uparrow\uparrow}(k+\tfrac{q}{2}, 
p-\tfrac{q}{2}, p+\tfrac{q}{2}, k-\tfrac{q}{2}) = &M^\Lambda_{k p}(q) + 
C^\Lambda_{k p}(q) - M^\Lambda_{\frac{k+p-q}{2}, \frac{k+p+q}{2}}(p-k) - 
C^\Lambda_{\frac{k+p-q}{2}, \frac{k+p+q}{2}}(p-k)\\
&+ P^\Lambda_{\frac{k-p+q}{2}, \frac{k-p-q}{2}}(k+p) - 
P^\Lambda_{\frac{k-p+q}{2}, -\frac{k-p-q}{2}}(k+p),
\label{eq:app:GammaPHuuuu}
\end{split}\\
\Gamma^{(4)\,\Lambda}_{\uparrow\downarrow\downarrow\uparrow}(k+\tfrac{q}{2}, 
p-\tfrac{q}{2}, p+\tfrac{q}{2}, k-\tfrac{q}{2}) = U + C^\Lambda_{k p}(q) - 
M^\Lambda_{k p}(q) - 2 M^\Lambda_{\frac{k+p-q}{2}, \frac{k+p+q}{2}}(p-k) + 
P^\Lambda_{\frac{k-p+q}{2}, \frac{k-p-q}{2}}(p+k).
\label{eq:app:GammaPHuddu}
\end{gather}
Symmetries allow to rewrite 
$\Gamma^{(4)\,\Lambda}_{\downarrow\uparrow\uparrow\downarrow}(k'-\tfrac{q}{2}, 
p + \tfrac{q}{2}, p - \tfrac{q}{2}, k'+\tfrac{q}{2}) = 
\Gamma^{(4)\,\Lambda}_{\uparrow\downarrow\downarrow\uparrow}(k'+\tfrac{q}{2}, 
p - \tfrac{q}{2}, p + \tfrac{q}{2}, k'-\tfrac{q}{2})$ and 
$\Gamma^{(4)\,\Lambda}_{\downarrow\downarrow\downarrow\downarrow}(k'-\tfrac{q}{2
}, p+\tfrac{q}{2}, p-\tfrac{q}{2}, k'+\tfrac{q}{2}) = 
\Gamma^{(4)\,\Lambda}_{\uparrow\uparrow\uparrow\uparrow}(k'+\tfrac{q}{2}, 
p-\tfrac{q}{2}, p+\tfrac{q}{2}, k'-\tfrac{q}{2})$. Note that the contributions 
of the particle-particle channel in Eq.~\eqref{eq:app:GammaPHuuuu} vanish in 
case only singlet pairing fluctuations are considered.

The flow of the effective interaction in the particle-particle channel is 
determined by the flow equation
\begin{equation}
\begin{split}
	\frac{d}{d\Lambda}P^\Lambda_{k k'}(q) = -\frac{1}{2} \intdrei{p}&
\frac{d}{d\Lambda}\bigl(G^\Lambda(\tfrac{q}{2}-p) 
G^\Lambda(\tfrac{q}{2}+p)\bigr) 
\Bigl(\Gamma^{(4)\,\Lambda}_{\uparrow\downarrow\downarrow\uparrow}(k+\tfrac{q}{2
}, \tfrac{q}{2}-k, \tfrac{q}{2}-p, \tfrac{q}{2}+p) 
\Gamma^{(4)\,\Lambda}_{\uparrow\downarrow\downarrow\uparrow}(\tfrac{q}{2}+p,
\tfrac{q}{2}-p,\tfrac{q}{2}-k',\tfrac{q}{2}+k')\\
& + 
\Gamma^{(4)\,\Lambda}_{\uparrow\downarrow\uparrow\downarrow}(\tfrac{q}{2}+k,
\tfrac{q}{2}-k,\tfrac{q}{2}-p,\tfrac{q}{2}+p) 
\Gamma^{(4)\,\Lambda}_{\downarrow\uparrow\downarrow\uparrow}(\tfrac{q}{2}+p,
\tfrac{q}{2}-p,\tfrac{q}{2}-k',\tfrac{q}{2}+k')\Bigr),
\end{split}
\end{equation}
where
\begin{gather}
\begin{split}
\Gamma^{(4)\,\Lambda}_{\uparrow\downarrow\downarrow\uparrow}(\tfrac{q}{2}+k,
\tfrac{q}{2}-k,\tfrac{q}{2}-p,\tfrac{q}{2}+p) = &U + P^\Lambda_{kp}(q) - 2 
M^\Lambda_{\frac{p-k+q}{2},\frac{k-p+q}{2}}(-k-p) + 
C^\Lambda_{\frac{k+p+q}{2},\frac{q-k-p}{2}}(k-p) \\
	& - M^\Lambda_{\frac{k+p+q}{2},\frac{q-k-p}{2}}(k-p),
\end{split}\\
\begin{split}
\Gamma^{(4)\,\Lambda}_{\uparrow\downarrow\uparrow\downarrow}(\tfrac{q}{2}+k,
\tfrac{q}{2}-k,\tfrac{q}{2}-p,\tfrac{q}{2}+p) = &-U - P^\Lambda_{k,-p}(q) + 2 
M^\Lambda_{\frac{k+p+q}{2},\frac{q-k-p}{2}}(k-p) - 
C^\Lambda_{\frac{p-k+q}{2},\frac{q+k-p}{2}}(-k-p) \\
	& + M^\Lambda_{\frac{p-k+q}{2},\frac{q+k-p}{2}}(-k-p)
\end{split}
\end{gather}
Note that due to symmetries, 
$\Gamma^{(4)\,\Lambda}_{\uparrow\downarrow\downarrow\uparrow}(\tfrac{q}{2}+p,
\tfrac{q}{2}-p,\tfrac{q}{2}-k',\tfrac{q}{2}+k') = 
\Gamma^{(4)\,\Lambda}_{\uparrow\downarrow\downarrow\uparrow}(\tfrac{q}{2}+k',
\tfrac{q}{2}-k',\tfrac{q}{2}-p,\tfrac{q}{2}+p)$ and 
$\Gamma^{(4)\,\Lambda}_{\downarrow\uparrow\downarrow\uparrow}(\tfrac{q}{2}+p,
\tfrac{q}{2}-p,\tfrac{q}{2}-k',\tfrac{q}{2}+k') = 
\Gamma^{(4)\,\Lambda}_{\uparrow\downarrow\uparrow\downarrow}(\tfrac{q}{2}+k',
\tfrac{q}{2}-k',\tfrac{q}{2}-p,\tfrac{q}{2}+p)$. As we neglect triplet pairing 
fluctuations, it is useful to work with a flow equation for the effective 
interaction in the singlet particle-particle channel, which is given by
\begin{equation}
\begin{split}
\frac{d}{d\Lambda}& P^{S,\Lambda}_{k k'}(q) =
\frac{1}{2}\Bigl(\frac{d}{d\Lambda} P^\Lambda_{k k'}(q) + \frac{d}{d\Lambda} 
P^\Lambda_{k, -k'}(q)\Bigr)\\
	= & -\frac{1}{4} \intdrei{p} 
\frac{d}{d\Lambda}\bigl(G^\Lambda(\tfrac{q}{2}-p)G^\Lambda(\tfrac{q}{2}
+p)\bigr)\Bigl(\Gamma^{(4)\,\Lambda}_{\uparrow\downarrow\downarrow\uparrow}
(\tfrac{q}{2}+k,\tfrac{q}{2}-k,\tfrac{q}{2}-p,\tfrac{q}{2}+p) - 
\Gamma^{(4)\,\Lambda}_{\uparrow\downarrow\uparrow\downarrow}(\tfrac{q}{2}+k,
\tfrac{q}{2}-k,\tfrac{q}{2}-p,\tfrac{q}{2}+p)\Bigr)\\
	&\quad 
\times\Bigl(\Gamma^{(4)\,\Lambda}_{\uparrow\downarrow\downarrow\uparrow}(\tfrac{
q}{2} +p,\tfrac{q}{2}-p,\tfrac{q}{2}-k',\tfrac{q}{2}+k') - 
\Gamma^{(4)\,\Lambda}_{\uparrow\downarrow\uparrow\downarrow}(\tfrac{q}{2}+p,
\tfrac{q}{2}-p,\tfrac{q}{2}-k',\tfrac{q}{2}+k')\Bigr).
\end{split}
\end{equation}

These flow equations were evaluated within the approximation scheme described 
in Sec.~\ref{sec:Approx} and Appendix~\ref{sec:NumDetailsAppendix}. As we do 
not renormalize the fermion-boson vertices, we set the external fermionic 
frequencies to zero, $k_0 = k_0' = 0$. In order to obtain flow equations for the 
bosonic exchange propagators, we project the flow equations for the coupling 
functions by averaging external fermionic momenta $\bs k$ and $\bs k'$ over the 
Fermi surface, as described in Ref.~\onlinecite{Eberlein2013}.

The flow equation for the self-energy~\eqref{eq:SelfEnergy} reads
\begin{equation}
\frac{d}{d\Lambda}\Sigma^\Lambda(k) = \intdrei{p} 
\bigl(\Gamma^{(4)\,\Lambda}_{\uparrow\uparrow\uparrow\uparrow}(k,p,p,k) + 
\Gamma^{(4)\,\Lambda}_{\uparrow\downarrow\downarrow\uparrow}(k,p,p,k)\bigr)
S^\Lambda(p),
\end{equation}
where the relevant components of the vertex are given by 
Eqs.~\eqref{eq:app:GammaPHuuuu} and~\eqref{eq:app:GammaPHuddu} with $q = 0$.
\end{widetext}

\section{Some details of numerical implementation}
\label{sec:NumDetailsAppendix}
In the dynamic approximation, the exchange propagators depend on frequency 
and momentum $q = (q_0, \bs q)$. These dependences are discretized on a 
three-dimensional grid. The momentum dependence is described with two grids in 
polar coordinates around $\bs q = \bs 0$ and $\bs q = \bs \pi$, with an 
increased density of grid points around these momenta, similarly to the grid 
used in Ref.~\onlinecite{Husemann2009}. The angular dependence is discretized 
using 3 - 7 angles in the first octant of the Brillouin zone, which is 
sufficient due to lattice symmetries, and 20 - 40 points for the radial 
dependence. The number of grid points was chosen higher in channels in which 
the effective interaction became large at low scales (for example the magnetic 
channel with momenta close to $\bs q \approx \bs \pi$ or the $d$-wave pairing 
channel with momenta close to $\bs q \approx \bs 0$), while less points were 
used for effective interactions that remained small during the flow (for 
example $s$-wave charge density wave fluctuations with $\bs q \approx \bs 0$). 
In case the leading instability was towards incommensurate states, we adjusted 
the momentum grid in such a way that the density of grid points is higher close 
to the anticipated position of the incommensurate peaks. The frequency 
dependence is discretized on a non-equidistant grid with typically 30 - 40 
frequencies between $q_{0,\text{min}} = 0$ and $q_{0,\text{max}} = 250$, where 
the density of grid points decreases towards higher frequencies. Linear 
interpolation is used for intermediate momenta and frequencies. We neglect the 
dependence of the effective interactions on the fermionic relative frequencies 
$k_0$, $k_0'$, which turned out to have a small impact on critical 
scales~\cite{Giering2012,Eberlein2013}. The flow equations are evaluated for 
$k_0 = k_0' = 0$, which allows to capture the ``bosonic'' features in the 
frequency dependence of the vertex~\cite{Rohringer2012} at small frequencies. 
In the static approximation, we neglect the dependence on $q_0$ and evaluate 
all effective interactions for $q_0 = 0$. The exchange propagators are then 
discretized on a two-dimensional grid and the numerical effort for solving the 
flow equations is significantly reduced.

These approximations transform the functional flow equations into a system of 
nonlinear ordinary differential equations. The coefficients on the right hand 
sides are given by three-dimensional integrals over loop momenta and frequencies 
in the dynamic approximation. In the static approximation all frequency 
integrals are solved analytically and only two-dimensional integrals have to be 
computed. The integrals are computed numerically with an adaptive algorithm with 
absolute and relative precision of $10^{-5}$ and $10^{-3}$, respectively. The 
flow equations were solved numerically with an adaptive fifth-order Runge-Kutta 
routine~\cite{Gough2009} with absolute and relative accuracy goals of 
$10^{-3}$. The numerical solution was started at a high scale $\Lambda_0 = 
100$, where the initial conditions can be computed in second-order perturbation 
theory.

%

\end{document}